\documentclass[aps,prl,reprint]{revtex4-1}
\usepackage[dvips]{graphicx}
\usepackage{amsmath}
\usepackage{color}

\begin{document}

\title{Third harmonic generation in graphene and few-layer graphite films}

\author{Nardeep Kumar$^1$, Jatinder Kumar$^1$, Chris Gerstenkorn$^1$, Rui Wang$^{1,2}$, Hsin-Ying Chiu$^1$, Arthur L. Smirl$^2$, and Hui Zhao$^1$}\email{huizhao@ku.edu}

\affiliation{$^1$Department of Physics and Astronomy, The University of Kansas, Lawrence, Kansas 66045, USA\\
$^2$Laboratory for Photonics and Quantum Electronics, The University of Iowa, Iowa City, Iowa 52242, USA}
\date{\today}

\begin{abstract}
We observe optical third harmonic generation from graphene and few-layer graphite flakes produced by exfoliation. The emission scales with the cube of the intensity of the incident near-infrared femtosecond pulses and has a wavelength that is one-third of the incident wavelength, both consistent with third harmonic generation. We extract an effective third-order susceptibility for graphene that is on the order of 10$^{-16}$ m$^2$/V$^2$, which is comparable to that for materials that are resonantly excited, but larger than for materials that are transparent at the fundamental and third harmonic wavelengths. By measuring a set of flakes with different numbers of atomic layers, we find that the emission scales with the square of the number of atomic layers, which suggests that the susceptibility of graphene is independent of layer number, at least for a few layers. 

\end{abstract}

\maketitle

Graphene is a monolayer of carbon atoms arranged in a hexagonal two-dimensional lattice. It has a linear dispersion relationship between energy, $E$, and wavenumber, $k$, $E(k) = \pm \hbar v_F k$, where the Fermi velocity $v_F \approx 10^6$~m/s [see Fig. 1(b)] \cite{pr71622,n438197,rmp81109}. Since the material became readily available less than a decade ago \cite{s306666}, it has been the subject of extensive investigations [{\it e.g.}, see Refs. \onlinecite{s3121191, nphoto4611} and references therein]. Graphene exhibits a number of unusual and remarkable transport properties that make it attractive for nano-electronic applications \cite{ieeeedl28282,l98206805,s327662}, including high mobilities \cite{s306666,s3121191,s327662,b82195414} and nearly-ballistic transport at room temperature \cite{nphoto4611,s3121191}; however, it is the optical properties of graphene that are of primary interest here.

The linear absorbance of graphene is flat and approximately 2.3\% across the entire visible spectrum \cite{s3201308,l101196405}. Thus, graphene can be considered to be both highly absorbing and/or highly transmitting, depending upon one's point of view or application. Since a negligible fraction ($<$ 0.1\%) is reflected \cite{l101196405}, 97.7\% of the incident light is transmitted. In this sense the sample is certainly transparent. On the other hand, if one were to assign an effective absorption coefficient, $\alpha$, to a monolayer of thickness 0.33 nm (even though it may be questionable to use macroscopic parameters, such as $\alpha$, for such thin samples), it would be very large ($\alpha \approx 7 \times 10^5$~/cm). As suggested by this large effective absorption coefficient, graphene interacts strongly with light. The strong broadband nature of the interaction of light with graphene is consistent with its linear bandstructure, where interband transitions of roughly equal strength are available throughout the visible. The combination of broadband transparency and the high mobilities mentioned earlier makes graphene a promising candidate for use as a transparent conductor \cite{n457706,nn5574,nl73394,acsnano2463} in photovoltaic devices \cite{nl8323,nl12133} and touch screens \cite{nn5574}. The high broadband absorption ({\it i.e.}, optical conductivity) and high carrier mobility suggest that graphene may find applications in a range of optically-controlled transport devices, such as broadband and ultrafast photodetectors \cite{nl91039,nn4839,nphoto4297,s334648,nl114134,nl112804,nn7114}, optical modulators \cite{n47464,nl121482}, plasmonic devices \cite{nn6630,nl113370,nl111814,s3321291,nphoto6259,nn7161}, broadband polarizers \cite{nphoto5411}, and magneto-optical devices \cite{np748}.

The strong broadband coupling of light with graphene also suggests that it might be an attractive nonlinear material for large bandwidth photonic applications. However, second order nonlinearities in graphene are expected to be weak because of symmetry considerations. For example, graphene mounted on a substrate has $C_{6v}$ symmetry, and therefore, the only allowed sources of second order responses are the interfaces \cite{bookboyd}. By comparison, for multilayer graphite, the bulk can contribute, depending upon the stacking arrangement \cite{apl95261910}. Indeed, second harmonic generation has been investigated both theoretically \cite{jetpl93366,b84045432,nl122032} and experimentally \cite{apl95261910,b82125411} in graphene and few-layer graphite films. Graphene flakes mounted on oxidized Si substrates have produced second harmonic signals, but the signal from the graphene is weaker than that from the substrate \cite{apl95261910,b82125411}. For bilayers and few-layer graphite samples similarly mounted, the second harmonic signal is dominated by the graphite films, but it is weak \cite{apl95261910,b82125411}.

In addition, the third order nonlinear optical responses associated with saturable absorption of graphene \cite{b83153410,b83121404,nl84248}, few-layer graphite films \cite{b82195414,b83121404,nl84248,nl84248,apl92042116,l101157402,apl94172102,b80245415,l104136802,apl97163103,nl101308,apl96081917,apl96173106,jap109084322,acsnano53278,nl111540}, and suspensions of graphene flakes \cite{cpl523104} have been measured and time-resolved using a variety of pump-probe configurations. Typically, in these experiments, the interband absorption of the pump pulse produces nonequilibrium electron and hole populations, which subsequently relax by carrier-carrier scattering, phonon emission, diffusion and recombination. The time-delayed probe pulse is used to measure the dynamic change in the absorption coefficient caused by Pauli blocking of the final and initial states by the electron and hole distribution functions, respectively. Consequently, the emphasis in these studies has been the determination of the carrier, phonon and recombination dynamics. Because the measured nonlinearities in such experiments are dynamic, cumulative, and incoherent, they are not well suited to defining and quantifying a third-order susceptibility {\it per se}, and little attention has been given to quantifying the magnitude of these nonlinearities.

By comparison, measurements of coherent third order nonlinearities are relatively rare \cite{nl112622,nl101293,l105097401}. Two-photon absorption has been measured in bilayer graphite flakes \cite{nl112622}, and a third order nonlinear susceptibility with magnitude, $|\chi^{(3)}|$, on the order of $10^{-16}$~m$^2$/V$^2$ was extracted, which is consistent with the quantum perturbation theory used by the authors. By comparison, in the same study  \cite{nl112622}, no two-photon absorption was observed in monolayer graphene, also consistent with the authorÕs theory which predicted a $|\chi^{(3)}| \approx 10^{-18}$~m$^2$/V$^2$. In contrast, however, ballistic photocurrents have been injected into multilayer epitaxial graphene samples by quantum interference between two-photon absorption and one-photon absorption, which suggests that the strength of the two-photon absorption is significant in graphene \cite{nl101293}. Moreover, $|\chi^{(3)}|$'s as large as 10$^{-15}$~m$^2$/V$^2$ have been reported for coherent four-wave mixing experiments in graphene \cite{l105097401}, and a study on coherent self-phase modulation in dispersion of graphene sheets containing effectively about 100 layers deduced a monolayer $|\chi^{(3)}|$ of the same order of magnitude \cite{nl115159}.  

\begin{figure}
\centering
\includegraphics[width=8.5cm]{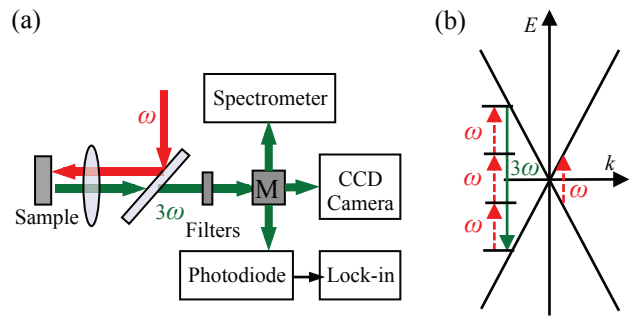}
\caption{(a) Experimental schematic: A fundamental ($\omega$, red) pulse is normally incident on the sample. The third harmonic (3$\omega$, green) is detected in the reflected direction by a CCD camera, a spectrometer or a photodiode connected to a lock-in amplifier. (b) Schematic diagram showing the linear dispersion of graphene, indicating the one-photon absorptions of $\omega$ and 3$\omega$ and illustrating the energy level description of third harmonic generation.}
\label{fig:setup}
\end{figure}

Here we report observation of third harmonic generation (THG) from graphene and few-layer graphite. The samples studied are graphene and few-layer graphite flakes produced by mechanical exfoliation and mounted on a 300-nm silicon dioxide layer grown on a 500-$\mu$m-thick silicon substrate. Flake thicknesses are determined by their optical contrasts \cite{apl91063124} and atomic force microscopy. Third harmonic emission is measured using the experimental geometry schematically sketched in Fig. \ref{fig:setup}(a). The fundamental near-infrared pulses ($\omega$, red) with an average power in the range of 1~-~10~mW are obtained from an optical parametric oscillator pumped by a Ti:sapphire laser at a repetition rate of 80 MHz. Each pulse is tightly focused onto the sample by a microscope objective, and the third harmonic generated by the sample ($3\omega$, green) is collected in the reflected direction by the same lens. Color filters are used to block the reflected $\omega$ beam. An adjustable mirror (M) is used to direct the $3\omega$ beam either to an imaging charge-coupled device (CCD) camera, a spectrometer, or a silicon photodiode connected to a lock-in amplifier. The camera is used to rapidly locate a flake having the desired thickness and to spatially image the 3$\omega$ spots. The spectrometer, which is equipped with a thermoelectric cooled silicon CCD camera, allows the detection of weak 3$\omega$ spectra. The combination of the photodiode, which has an internal amplifier, and the lock-in amplifier provides for the continuous monitoring of the 3$\omega$ power as experimental parameters are varied.

\begin{figure}
\centering
\includegraphics[width=8.5cm]{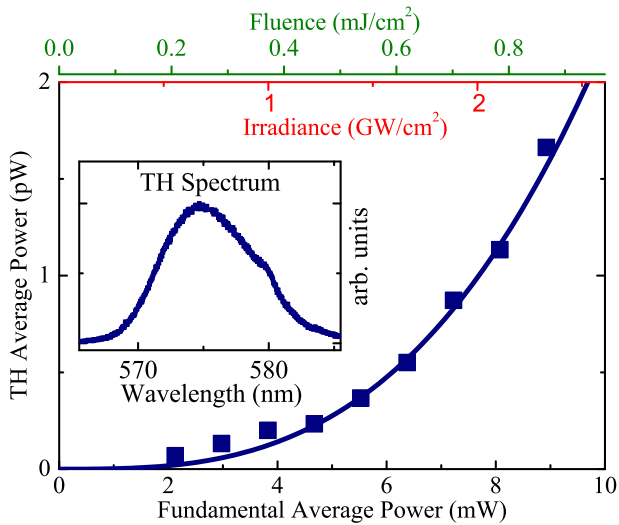}
\caption{The average power of the third harmonic as a function of the average power of the incident fundamental (solid squares), and the spectrum of the third harmonic (solid squares, inset). The solid line is a cubic fit ($\chi^{(3)} = 4 \times 10^{-17}$ m$^2$/V$^2$) of Eq. (1) to the data using parameters appropriate for our experimental conditions, as discussed in the text. }
\label{fig:monolayer}
\end{figure}

The 3$\omega$ signal produced by a graphene flake when it is irradiated by an $\omega$ beam with horizontal linear polarization is summarized in Fig.~\ref{fig:monolayer}. The spectrum of the emitted 3$\omega$ light produced by an $\omega$ beam with an average power of 10 mW is shown in the inset of Fig.~\ref{fig:monolayer}. The central wavelength of 575.5 nm is within 0.4\% of the separately measured $\omega$ wavelength (1720.4 nm) divided by 3 ({\it i.e.}, 573.5 nm), which confirms that we are measuring the third harmonic [see Fig.~\ref{fig:setup}(b) for the energy diagram of THG]. When the laser spot is moved from the flake to the bare substrate, no 3$\omega$ signal is observed under the same conditions. Also, the second-harmonic signals from the graphene flakes, the Si/SiO$_2$ substrate, and their interfaces, which were previously observed in similarly mounted samples \cite{apl95261910,b82125411}, are much smaller than the third harmonic and are not measurable under our conditions. In addition, by using an analyzing polarizer in front of the spectrometer, we find that the emitted 3$\omega$ radiation is linearly polarized along the same direction as the $\omega$ pulse.

The main panel of Fig.~\ref{fig:monolayer} shows the 3$\omega$ average power as a function of the $\omega$ average power. For convenience, the corresponding $\omega$ peak on-axis irradiance and pulse fluence are also shown on the top axis. The average 3$\omega$ power is calibrated by comparing the measured 3$\omega$ signal on the spectrometer to that produced by a laser pulse of known power at 560 nm incident on the substrate. In this way, the reflection of the substrate, the loss of all optics, and the sensitivity of the spectrometer are included in the detector calibration. The measured signal is consistent with a cubic dependence on incident $\omega$ power, as indicated by the comparison with solid line in the main panel of Fig.~\ref{fig:monolayer} and as expected for third harmonic signal. The small deviation from cubic behavior at extremely low power levels can be attributed to unremoved noise background in the signal.

In an attempt to quantify the strength of the third harmonic signal produced by graphene, we now extract an effective third-order susceptibility, $\chi^{(3)}$, by comparing the 3$\omega$ signal shown in Fig.~\ref{fig:monolayer} with the well-known result for a thin bulk sample in the limit of an undepleted fundamental \cite{bookboyd}:
\begin{equation}
\mathcal{E}_{3\omega}=\frac{1}{4}\frac{i3\omega}{2n_{3\omega}c}d\chi^{(3)}\mathcal{E}_\omega^3,
\end{equation}
where $c$ is the speed of light in a vacuum, $d$ is the sample thickness, $n_{3\omega}$ is the index of refraction of the material at 3$\omega$, and $\mathcal{E}_{3\omega}$ ($\mathcal{E}_{\omega}$) is the magnitude of the 3$\omega$ ($\omega$) field and where each field is related to its irradiance by $I_i = n_i \epsilon_0 c \mathcal{E}_i \mathcal{E}_i^* / 2$. Furthermore, if the irradiance associated with each fundamental pulse is taken to be Gaussian in space and time,
\begin{equation}
I_\omega(r,t) = I_0 e^{-4\mathrm{ln}2(r/W)^2}e^{-4\mathrm{ln}2(t/\tau)^2},
\end{equation}
with a diameter $W$ and a pulsewidth $\tau$ (both in full width at half maxima), the peak on-axis irradiance, $I_0$, is related to the average power, $\bar{P_\omega}$, by
\begin{equation}
\bar{P_\omega} = f \frac{\sqrt{\pi} \tau}{\sqrt{4\mathrm{ln}2}} \frac{\pi W^2}{4\mathrm{ln}2}I_0,
\end{equation}
where $f$ is the repetition rate of the laser. While the use of bulk concepts, such as susceptibilities and indices of refraction, to describe a single atomic layer is questionable, this procedure allows us to assign an effective material parameter that allows comparison of THG in graphene to that in other well-known bulk materials. Using values appropriate to our experiment [i.e., $d$ = 0.33 nm, $n_\omega \approx n_{3\omega}$ = 2.4, $W$ = 3.5 $\mu$m, $\tau$ = 320 fs, $\lambda$ =1720.4 nm, and $f$ = 81 MHz], these expressions are used to extract a value of $|\chi^{(3)}| \sim 4 \times 10^{-17}$~m$^2$/V$^2$ from Fig.~\ref{fig:monolayer}. We note that the sample used for these measurements is shaped like a rectangle with a width $\sim 1~\mu$m and a length much larger than the focused laser spot diameter. Consequently, the overlap between fundamental focused laser spot and the sample is imperfect. The fit shown in Fig.~\ref{fig:monolayer} is corrected for this incomplete overlap. The fit shown in Fig.~\ref{fig:monolayer} is not corrected for multiple reflections associated with the layered oxide-Si structure.  A calculation based on light propagation through our specific layered structure,\cite{josab222144,b82125411} which incorporates third harmonic polarizations, indicates that the $\chi^{(3)}$ value extracted in this case is changed only by roughly a factor of two from that reported above, which is within our experimental uncertainty.  Therefore, for pedagogical simplification, we ignore those complications here. We also note that the upper limit on the fluence (and irradiance) shown in Fig.~\ref{fig:monolayer} is determined by damage to the sample.

\begin{figure}
\centering
\includegraphics[width=8.5cm]{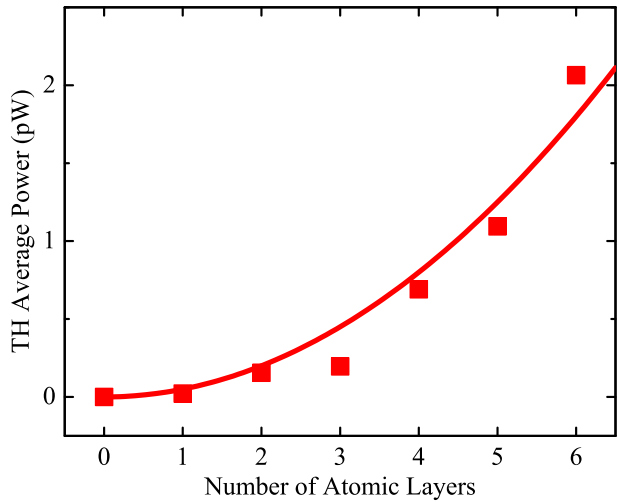}
\caption{The average power of the third harmonic as a function of the number of atomic layers for an average fundamental power of 1 mW (solid squares). The solid line is a quadratic fit ($\chi^{(3)} = 1 \times 10^{-16}$ m$^2$/V$^2$) of Eq. (1) to the data using parameters appropriate for our experimental conditions, as discussed in the text. }
\label{fig:multilayer}
\end{figure}

We also investigate the nonlinear optical response as the sample evolves from monolayer graphene to multi-layer graphite by measuring the third harmonic from flakes of thickesses ranging from 1 to 6 atomic layers mounted on a single Si/SiO$_2$ substrate for an average power of 1 mW, as shown in Fig.~\ref{fig:multilayer}. The data point labeled as having zero atomic layers represents the response of the bare substrate, and it confirms that any third harmonic generated by the substrate is negligible on our measurement scales. The solid curve in Fig.~\ref{fig:multilayer} is a fit of Eq. 1 to the data for an effective third-order susceptibility of $|\chi^{(3)}| \sim 1 \times 10^{-16}$~m$^2$/V$^2$, and it has a quadratic dependence on sample thickness. Notice that this value is $\sim$ 2.5 times larger than that extracted from Fig.~\ref{fig:monolayer}. To the degree possible, the experimental conditions are held constant for the two sets of measurements; however, they are performed on different days, on different flakes, of different sizes, mounted on different substrates. Given the cubic dependence of the third order interaction on the field and given the inherent uncertainties and variation in the experimental parameters, this represents reasonable agreement between the two measurements. For these reasons, $|\chi^{(3)}| \sim 10^{-16}$~m$^2$/V$^2$ should be regarded as an order of magnitude estimate of the susceptibility. This magnitude is one order of magnitude smaller than those estimated from recent four-wave mixing \cite{l105097401} and self-phase modulation \cite{nl115159} experiments, but more than two orders of magnitude larger than that from two-photon absorption experiment and quantum perturbation calculation \cite{nl112622}. Further efforts, from both theoretical and experimental sides, are needed to fully understand third order response of graphene. 

The deduce effective $\chi^{(3)}$ is several orders of magnitude larger than that for transparent materials  \cite{bookboyd}. This is because third order response is dramatically enhanced by the resonant nature of the intermediate and final transitions needed for the THG \cite{bookboyd}. The nonlinear response of graphene is also broadband, owing to its bandstructure. On the other hand, however, $\chi^{(3)}$ of graphene is comparable to those associated with resonant transitions in other materials ({\it e.g.} metal nanoparticles) \cite{bookboyd}. The presence of resonant transitions at $\omega$ and 3$\omega$ in graphene not only enhances $\chi^{(3)}$, but also ensures strong linear (and nonlinear) absorption at both frequencies. This absorption is a detriment to the third order processes, because it reduces the fundamental intensity needed for THG and leads to the subsequent absorption of the third harmonic \cite{pe44924}. In fact, one can define a figure of merit (FOM) for the THG as the ratio of the total number of third harmonic photons generated to the total number of photons ($\omega$ and 3$\omega$) absorbed. It is straightforward to show that this FOM is, in turn, proportional to the ratio $\chi^{(3)} / \alpha$. Thus, while graphene and thin graphite films (and indeed all materials with resonant transitions) have large $\chi^{(3)}$'s, they do not have unusually large FOMÕs. For example, for our experimental conditions, a pulse with a fluence of ~100 $\mu$J/cm$^2$ (peak intensity of $\sim$ 0.3 GW/cm$^2$) would have a third harmonic conversion efficiency of less than 10$^{-10}$, but would lose 0.023 of its energy to linear absorption.

In summary we have observed third harmonic generation from graphene and few-layer graphite films. For a few layers, the emission scales with the cube of the intensity and the square of the number of atomic layers. It is characterized by an effective third-order susceptibility of the order of $\chi^{(3)}(3\omega;\omega,\omega,\omega) \sim 10^{-16}$~m$^2$/V$^2$. Such a strong third order response originates from the resonant nature of the light-graphene interaction, which also causes strong absorption of light. 

This work was support in part by the US National Science Foundation under Awards No. DMR-0954486 and No. EPS-0903806, and matching support from the State of Kansas through Kansas Technology Enterprise Corporation. 


%

\end{document}